
\magnification =  1200
\nopagenumbers
\headline{\ifnum\pageno=1 \nopagenumbers
\else \hss\number \pageno \fi}

\overfullrule=0pt
\newdimen\nude\newbox\chek
\def\slash#1{\setbox\chek=\hbox{$#1$}\nude=\wd\chek#1{\kern-\nude/}}
\def\QS{\slash{Q}}
\def\wpp{\bf{P}}
\centerline {\bf } \par \bigskip \vskip 1cm
\raggedbottom
\baselineskip 15 pt
\def\cit#1#2#3{{\bf{#1}} (#2) #3}
\vskip 0.50 truecm

\def\section#1#2{\medskip\noindent{\bf#1.\quad
#2}\par\smallskip}
\def\abstract#1{\vskip 15 pt\midinsert\narrower
\smallskip
 \noindent{\bf ABSTRACT}\quad{#1}
\smallskip
\endinsert}

\def\cen{\centerline}

\smallskip
\def\section#1#2{\medskip\noindent{\bf#1.\quad
  #2}\par\smallskip}
\def\cen{\centerline}

\cen{{\bf SOFT PHOTON PRODUCTION RATE  IN RESUMMED }}

\cen{{\bf PERTURBATION THEORY OF HIGH TEMPERATURE QCD}}
\vskip 10 pt
\vskip 20 pt
\centerline
{\bf R.~Baier~$^1$, S.~Peign$\bf {\acute e}$~$^2$ and D.~Schiff~$^2$}
\vskip 10 pt
\centerline{{\it $^{1 }$Fakult\"at f\"ur Physik,
Universit\"at Bielefeld, D-33501 Bielefeld, Germany}}
\centerline{{\it $^{2 }$LPTHE{$^{\dagger}$}, Universit\'e
Paris-Sud, B\^atiment 211, F-91405 Orsay, France}}
\vskip 10pt
\noindent
\footnote{} {$^{\dagger}$ Laboratoire associ\'e du
Centre National
de la Recherche Scientifique}
\def\ref#1{[#1]}
\vskip 25 pt
\noindent
\abstract{We calculate the production rate of soft real photons
from a hot quark -- gluon plasma using Braaten -- Pisarski's
perturbative resummation method. To leading order in the QCD
 coupling constant $g$ we find a logarithmically divergent result
for photon energies of order $gT$, where $T$ is the plasma
temperature. This divergent behaviour is due to unscreened mass
singularities in the effective hard thermal loop vertices
in the case of a massless external photon.}
\vskip 40 pt
\noindent
LPTHE-Orsay 93/46
\par\noindent
BI-TP 93/55
\par\noindent
November 1993
\vskip 20 pt
\eject
\section{1}{INTRODUCTION}

\vskip 5 mm
Theoretical investigations predict the formation of a quark - gluon
plasma (QGP)
 in high-energy
heavy ion collisions. Many signatures for this new phase of nuclear matter have
been proposed, in particular electromagnetic ones : photon
production [1] by the QGP is expected to
 be an interesting signal, as the mean free path of
the photon $\gamma$ in the thermal medium is expected to be larger than the
size of the
plasma, at least when  the energy of the  $\gamma$  is not too small. \par
 The present paper is
concerned with real direct photon production, assuming that the photon
is not thermalized.
 The production rate of hard photons (with
energy $E \sim T$, where $T$ is the plasma temperature)
 has already been studied in great detail [1 - 7].
Especially when applying
  the framework of the resummed
perturbative expansion of Braaten and Pisarski [8 - 11],
 it has been
demonstrated that mass singularities due to the exchange of massless quarks are
shielded  by effects due to Landau damping [5 - 6].
 In the following  we are dealing with the
soft photon production rate ($E \sim gT$, where $g$ is the QCD coupling
constant),
whose calculation requires not only to use resummed quark propagators but also
dressed vertices. This should allow us to extract the (finite) leading
contribution to the production rate, thus completing the list of
predictions of electromagnetic signals:
 real  and virtual [12 - 14] photon rates.
 \par
The main result of the present work is that contrary to the hard photon
case, for soft real photons the
resummation advocated by Braaten and Pisarski does not succeed to screen mass
singularities, i.e. in this case the resummed perturbative
expansion fails to give a finite contribution  at leading order for a physical
quantity. \par
The outline of the paper is the following : in section 2, we briefly review
the situation for hard photon production. In section 3, we deal with soft
photons, showing the origin of mass singular terms and exhibiting them,
then extracting the singular contribution using dimensional
regularisation. Section 4 is devoted to a short discussion.
\par
\vskip 5 mm
\section{2}{HARD PHOTON PRODUCTION RATE}
\vskip 5 mm
The Born calculation of the hard photon production rate, which
uses a bare internal quark propagator for the first order annihilation and
Compton scattering amplitudes, gives as the leading term [2 - 7]
in the limit of vanishing bare quark mass $m$:
\vskip 2 mm
$$E {dW \over d^3 \vec{p}} \simeq {e_q^2 \alpha \alpha_s \over 2 \pi^2} T^2
e^{-E/T} \ell n \left ( { ET \over m^2} \right ) \ \ \ , \eqno(1)$$
\vskip 3 mm
\noindent where $E > T$.
$\alpha$ is the fine structure constant, $e_q$ the quark charge and
$\alpha_s = g^2 / 4 \pi$.  \par
The obvious problem with eq. (1) is that the photon production rate is
divergent
when $m$ tends to zero.
 Thus the leading  order Born calculation is not satisfactory.
\par
Indeed, when the  momentum transfer of the exchanged quark, e.g.
in the Compton process,
 is soft of $O( gT )$
one has to resum an infinite set of diagrams contributing
to the same order as the Born term. This resummation program has been proposed
by
Braaten and Pisarski [8 - 11]. In case of the propagator it amounts
 to replace the
bare  by an effective one, whenever the momentum is soft. \par
The main characteristic property of the effective propagator is that it is
dynamically screened on the momentum scale
of  order $gT$ for space-like momenta, due to the mechanism
of Landau damping. The soft scale is characterized  by the
fermion mass induced by temperature,
 i.e.  by $m_f = \sqrt{2 \pi \alpha_s /3}~ T$.
It acts as an infra-red cut-off, and  in eq. (1) one may replace
$m$  by $m_f$ [3].
 This heuristic
manipulation yields in fact the right result in the leading-logarithm
approximation. The rigorous calculation [5 - 6], which also allows to find the
numerical
coefficient appearing inside the log, leads to:
\vskip 2 mm
$$E {dW \over d^3 \vec{p}} \simeq {e_q^2 \alpha \alpha_s \over 2 \pi^2} T^2 \
e^{-E/T}
\ell n \left ( {c \over \alpha_s} {E \over T} \right ) \eqno(2)$$
\vskip 3mm
\noindent with $c \simeq 0.23$, when $E > T$. \par
One obtains a finite production
rate for  hard photons. The $\ell n \left ( {1 \over \alpha_s} \right)$
dependence in eq. (2) is a reminiscence of the logarithmic divergence of the
Born term, which indeed becomes dynamically screened after resummation.

In the following we
want to know whether the resummed perturbative expansion achieves
the same screening of mass singularities in the case of soft photon
 production, i.e. for photon energies of $O (gT)$.
\par
\vskip 5 mm
\section{3}{SOFT PHOTON PRODUCTION RATE}
\vskip 5 mm
The production rate may be computed in a systematic way by evaluating the
imaginary part of the photon polarization tensor:
\vskip 2 mm
$$E {dW \over d^3 \vec{p}} = - {1 \over (2 \pi)^3} n_B(E) \ Im \
\Pi_{\mu}^{\mu}(E,
\vec{p}) \ \ \ , \eqno(3)$$
\vskip 3 mm
\noindent where $\Pi_{\mu}^{\mu}$  is first calculated in the euclidean
formalism\footnote{*}{$\Pi_{\mu}^{\mu}$ is evaluated for $p_4 = 2 \pi nT$ and
then continued according to $ip_4 \to E$. In the imaginary time formalism, the
euclidean Dirac algebra $\{ \gamma^{\mu} , \gamma^{\nu} \} = 2 \delta^{\mu
\nu}$ is
used.}. The Bose - Einstein distribution is denoted by $n_B$.
 \par
 When the photon energy $E$ is of order $gT$, either $k$ and $k'$
(Fig. 1) are soft ($\sim gT$), and both quark propagators have to be
resummed, or $k$ and $k'$ are hard ($\sim T$), but the latter contribution is
suppressed by a factor $g^2$ and we shall neglect it. By evaluating the soft
photon production rate according to eq. (3)  we thus consider both internal
quark propagators as soft ones. As the photon momentum $P = k+ k'$ is soft,
vertices have also to be resummed and the relevant photon polarization tensor
entering eq. (3) is shown in Fig. 1. \par
 As the internal quark propagators are resummed, we expect
that screening occurs as for the hard photons, and no divergence appears when
$k$ or $k'$
are vanishing. However, the
introduction of effective vertices, though necessary to take into account all
diagrams contributing to the rate at
 leading order in $g$, will be shown to lead to
unscreened collinear divergences. \par
\vskip 5 mm
\noindent {\bf 3.1 Resummed photon self energy} \par
\vskip 5 mm
In order to evaluate the production rate (eq. (3)) we first consider
$\Pi_{\mu}^{\mu}$~:
\vskip 2 mm
$$\Pi_{\mu}^{\mu}(E, \vec{p}) = e_q^2 \ e^2 \ N_c \ T \sum_{k_4} \int {d^3
\vec{k}
\over (2 \pi)^3}
3~ tr \left [ ^{\ast} \Delta (k) ^{\ast} \Gamma^{\mu} (k, k'; -P)
^{\ast} \Delta(-k') ^{\ast}
\Gamma_{\mu}(-k', -k; P) \right ] \ \ \  , \eqno(4)$$
\vskip 2 mm
\noindent where $N_c$ is the number of colours
 and $^{\ast} \Delta (k)$ is the effective quark propagator~:
\vskip 2 mm
$$^{\ast} \Delta(k) = {1 \over 2} \left ( {\gamma \cdot k_+ \over {D_+(k)}} +
{\gamma \cdot k_- \over {D_-(k)}} \right ) \ , \eqno(5)$$
\noindent with
$$k_{\pm} = \left ( 1, \pm i \ \widehat{k} \right ) \ \ \ ,  \ \ \
 \widehat{k} = {\vec{k}/ \vert \vec k \vert }\ \ \ .
 $$
\vskip 3 mm
\noindent The functions $D_{\pm} (k)$  are given in [13 - 17].
\noindent The effective quark - photon vertex [10 - 11, 13 - 14] is represented
by:
\vskip 2 mm
$$^{\ast} \Gamma^{\mu} = \gamma^{\mu} + m_f^2 \int {d \Omega \over 4 \pi}
{Q^{\mu} {\QS} \over \left ( Qk \right ) \left ( Qk' \right ) } $$
$$Q = \left ( i, \widehat{Q} \right ) \ \ \ . \eqno(6)$$
\vskip 3 mm
 The second
term in the r.h.s. of eq. (6) is the hard thermal
 loop correction -- in terms of an angular integral --  to the bare
vertex $\gamma^{\mu}$. $Q$ is a light-like vector, $Q^2 = 0$;
 the inner product $Q \cdot k = Q_4 k_4 + \widehat{Q} \cdot \vec k$
is denoted by $(Q k)$. \par
The Dirac trace in eq. (4) is split into three terms according to the number
 of hard loop corrections, cf. eq. (6): they are
denoted by $tr(0)$, $tr(1)$, $tr(2)$,
respectively. We get~:
\vskip 3 mm
$$\Pi_{\mu}^{\mu} = e_q^2 \ e^2 \ N_c \ T \sum_{k_4} \int {d^3 \vec{k} \over (2
\pi)^3} \left [ tr(0) + tr (1) + tr (2) \right ] \ \ ,
\eqno(7)$$
\vskip 3 mm
where
$$tr(0) = 2 \sum_{i,j = \pm} ~ {{\left ( k_i k'_j \right )}
 \over{ D_i \ D'_j } } \ \ ,
 \eqno(7 a)$$
\vskip 3 mm
$$tr(1) =
      - 4 \ m_f^2 \int {d \Omega \over 4 \pi} \ {1 \over (kQ) \ (k'Q) }\
 \sum_{i,j = \pm} ~ {{\left ( k_i Q \right ) \left ( k'_j  Q \right )}
 \over{ D_i \ D'_j } } \ \ ,
  \eqno(7 b)$$
\vskip 3 mm
$$tr(2) = - m_f^4 \int {d \Omega_1 \over 4 \pi} \int {d \Omega_2 \over 4 \pi} \
{\left ( Q_1 \ Q_2 \right ) \over \left ( k \ Q_1 \right ) \left ( k \ Q_2
\right ) \left ( k' \ Q_1
\right ) \left ( k' \ Q_2 \right ) } \   \eqno(7 c)$$
$$ \cdot \
 \sum_{i,j = \pm} ~
  {1 \over D_i \ D'_j} \left [ \left ( Q_1 \ k_i \right ) \left ( Q_2
\ k'_j \right ) + \left ( Q_1 \ k'_j \right ) \left ( Q_2 k_i \right ) - \left
(
Q_1 \ Q_2 \right ) \left ( k_i \ k'_j \right ) \right ]   \ \   .$$
\vskip 3 mm
\noindent The primed quantities depend on the momentum $k' = P - k$. \par
 To obtain    the imaginary part of $\Pi_{\mu}^{\mu}$ we use
 the identity [13 - 14]~:
\vskip 2 mm
$$Im \ T \sum_{k_4=2 \pi n T} f \left ( i k_4 \right ) \ f' \left ( i \left (
p_4 - k_4 \right ) \right ) = $$
$$ \pi \left ( 1 - e^{E/T} \right ) \int_{-
\infty}^{\infty} d \omega
\int_{- \infty}^{\infty} d \omega ' n_F(\omega)
n_F(\omega ') \delta \left ( E - \omega - \omega ' \right ) \rho (\omega ) \rho
'
( \omega ' ) \ \ ,   \eqno(8) $$
\vskip 3 mm
\noindent where $n_F$ is the Fermi - Dirac distribution, and
 $\rho$, $\rho '$ are the spectral densities associated with $f$,
$f'$~, respectively:
\vskip 2 mm
$$\rho ( \omega ) = \lim_{\varepsilon \to 0} {1 \over \pi} \ Im \ f(\omega + i
\varepsilon ) \ \ \ .\eqno(9)$$
\vskip 3mm
\noindent Eq. (8) can be used only if the dependence of $\Pi_{\mu}^{\mu}$ on
$ik_4$ and $ik'_4$ is factorized. Thus, as proposed by Wong [14], it is
convenient to take the discontinuity of $\Pi_{\mu}^{\mu}$ in the
factorized form (7), before integrating over $d \Omega$. \par
 When the continuation
$ik_4 \to \omega + i \varepsilon$ is performed,
the functions  with non-vanishing discontinuities
 are $\left [ D_{\pm}(k) \right
]^{-1}$, $\left [ \left ( Qk \right ) D_{\pm}(k) \right ]^{-1}$, $\left [ \left
(
Q_1k \right ) \left( Q_2k \right ) D_{\pm}(k) \right ]^{-1}$ appearing in the
terms
$tr(0)$, $tr(1)$, $tr(2)$ of eq. (7). The associated spectral densities are
denoted by $\rho_{\pm}$, $\sigma_{\pm}$, $\tau_{\pm}$, respectively. For
space-like momentum $k$~, $\omega < \vert \vec{k}  \vert$, they read:
\vskip 2 mm
$$\rho_{\pm} =
 { 1 \over \pi}  Im \left [ D_{\pm} \right ] ^{-1} =
 \beta_{\pm} \left ( \omega , | \vec{k} | \right ) \quad ,
 \eqno(10 a)$$
$$\sigma_{\pm} =
 { 1 \over \pi}  Im
  \left [\left ( Qk \right ) D_{\pm}  \right
]^{-1} =
 {\wpp} \left ( {1 \over Qk} \right ) \beta_{\pm} - \delta (Qk)
\alpha_{\pm} \quad ,
 \eqno(10 b)$$
$$\tau_{\pm} =
 { 1 \over \pi}  Im
  \left [ \left ( Q_1k \right ) \left (
Q_2k \right ) D_{\pm} \right ]^{-1} =
 \left [ {\wpp} \left ( {1 \over Q_1k} \right ) {\wpp} \left ( {1 \over Q_2k}
\right ) - \pi^2 \delta \left ( Q_1k \right ) \delta \left ( Q_2k \right
) \right ] \beta_{\pm}$$
$$- \left [ {\wpp} \left ( {1 \over Q_1k} \right ) \delta \left ( Q_2k \right )
+ {\wpp} \left
( {1 \over Q_2k} \right ) \delta \left ( Q_1k \right
) \right ] \alpha_{\pm} \ \ \ ,
 \eqno(10 c)$$

\vskip 3 mm
\noindent where
$$ \lim_{\varepsilon \to 0}
   { 1 \over { D_{\pm}
(ik_{4} \to \omega + i \varepsilon )  }} =
 \alpha_{\pm} + i \pi \beta_{\pm}  \ \ .
 \eqno(10 d)$$
\vskip 3 mm
\noindent ${\wpp}$ denotes the principal part
prescription. The detailed expressions for the functions
$\alpha_{\pm}$ and $\beta_{\pm}$ can be found in [13 - 17].
 We
show below that the mass singularities  arise only when both $k$ and
$k'$ are space-like momenta.
\vskip 5 mm
\noindent {\bf 3.2 Mass singular contributions} \par
\vskip 5 mm
  Taking the imaginary part of $\Pi_{\mu}^{\mu}$ leads to products
of the type $\rho_{\pm} \rho_{\pm} '$,
 $\sigma_{\pm} \sigma_{\pm} '$ and $\tau_{\pm} \tau_{\pm} '$ in eqs. (7 a), (7
b) and
(7 c). The next step is to integrate over $d \Omega$, and over $d \Omega_1$, $d
\Omega_2$,
respectively. \par
 Only the $1 \over (Qk)$ factors can produce singularities:
e. g.  in the product of terms $\sigma_{\pm} \sigma_{\pm} '$ arising from eq.
(7 b) we have
\vskip 2 mm
$${\wpp} \left ( {1 \over Qk} \right ) \delta (Qk') = {\wpp} \left ( {1 \over
QP} \right )
\delta (Qk') = {1 \over (QP)} \delta (Qk') \  ,  \eqno (11)$$ \vskip 3 mm
\noindent (as $(QP) \geq 0$
 the ${\wpp}$ prescription is dropped); i.e.
when $(QP) = 0$, i.e. $\widehat{Q} \rightarrow - {\vec{p} \over E}$,  $(Q^2 =
P^2 = 0)$, a non-integrable
singularity appears.
In the following we regularize this singularity
by  using dimensional regularisation of the angular integral over $d \Omega$ in
$D =
3 + 2 \widehat{\varepsilon}$ dimensions, with $\widehat{\varepsilon} > 0$~,
but keeping only the singular parts~:
\vskip 2 mm
$$ \int {d \Omega \over 4 \pi}    \rightarrow
  ~\int {d \Omega \over 4 \pi} \vert_{reg}
 {\buildrel \rm def \over =} ~ {1 \over 2} \int_0^{\pi} d
\theta \ \sin^{D-2} \theta = {1 \over 2} \int_{-1}^1 d \cos \theta \left ( 1 -
\cos^2 \theta \right )^{\widehat{\varepsilon}} \ \ \ . \eqno(12)$$
\vskip 3 mm
By taking the discontinuity of eq. (7) using eq. (8), we thus retain only the
products ${\wpp} \left ( {1 \over Qk} \right ) \ \delta(Qk')$ in order to
compute
the leading (singular) contribution to the soft photon production rate.
The  singularity arises when $(Qk) = (Qk') = 0$, which is possible only
for space-like $k$ and $k'$. For this reason  we restrict ourselves to this
domain, since all other contributions are regular\footnote{*}{No singularity is
produced
by the ${\wpp} {\wpp}$ or $\delta \delta$ products
present in the terms proportional to
 $\sigma_{\pm} \sigma_{\pm} '$ and $\tau_{\pm} \tau_{\pm} '$
. }. \par
In some more detail we describe the procedure for the term
 $tr(1)$, eq. (7 b).
 The part giving rise to the singularity
 reads -- after regularisation according to eq. (12)~:
\vskip 2 mm
$$ \left . Im \ \Pi_{\mu}^{\mu} \right |_{1, reg} = - 4 \ m_f^2
 \int [dk]~ \int {d \Omega
\over 4 \pi} \vert_{reg}  \ {1 \over (QP)} \  \eqno(13)$$
$$ \cdot  \left \{ \delta (Qk')
 \left [ \beta_+ \left ( 1 + \widehat{k} \cdot
 \widehat{Q} \right )
+ \beta_- \left ( 1 - \widehat{k} \cdot \widehat{Q} \right ) \right ] \left [
\alpha '_+ \left ( 1 + \widehat{k'} \cdot \widehat{Q} \right ) + \alpha '_-
\left ( 1 - \widehat{k'} \cdot \widehat{Q} \right ) \right ] \right .$$
$$\left . + sym \left ( k \longleftrightarrow k' \right ) \right \}$$
\vskip 3 mm
\noindent where the continuation
 e.g. $(Qk) \Rightarrow  \omega + \widehat{Q} \cdot \vec{k}$
is implied.  $\alpha_{\pm},
 \beta_{\pm}$ are functions of $\omega$ and $k = | \vec{k} |$.
The integrations with respect to $k, \omega, \omega '$ are indicated by the
short-hand notation:
$$
\int [ dk ] \equiv e^2 \ e_q^2 \ N_c \ \pi(1 - e^{E/T})  \int {d^3 \vec{k}
\over (2 \pi)^3} \int_{-|k|}^{|k|} d \omega \int_{-|k'|}^{|k'|} d \omega ' \
n_F
(\omega) \ n_F (\omega ') \delta(E - \omega - \omega ')
 \ .\eqno(14)$$
In the limit  $\widehat{\varepsilon} \to 0$
$ \left . Im \ \Pi_{\mu}^{\mu} \right |_{1, reg} $ behaves as
 $1 / \widehat{\varepsilon}$: the residue is determined by
 replacing $\widehat{Q}$ by $- {\vec{p} \over E}$.
\noindent The  integral over $d \Omega$ is then computed for
$\widehat{\varepsilon} \to 0$~:
\vskip 2 mm
$$~ \int {d \Omega \over 4 \pi} \vert_{reg} {1 \over (QP)} \simeq {1
\over 2 E~ \widehat{\varepsilon}} \ \ \ . \eqno(15)$$
\vskip 3 mm
\noindent The leading  divergent behaviour expressed in terms of the factor ${1
/
\hat{\varepsilon}}$ finally becomes~:
\vskip 2 mm
$$\left . Im \ \Pi_{\mu}^{\mu} \right |_{1, reg} = - 2 \ m_f^2
          {1 \over
\widehat{\varepsilon}} ~\int [dk]~ \delta (Pk) \   \eqno(16)$$
$$  \cdot  \left \{ \left ( \beta_+ \left (
1 - {\omega \over k} \right ) + \beta_- \left ( 1 + {\omega \over k}
\right ) \right ) \left ( \alpha '_+ \left ( 1 - {\omega ' \over k'} \right
) + \alpha ' _- \left ( 1 + {\omega ' \over k'} \right ) \right ) \right .$$
$$\left. + sym (\ \omega, k \longleftrightarrow \ \omega ', k' ) \right \} \ \
\ .$$
\vskip 3 mm
 Next we discuss the contribution including two
 vertex corrections, i.e. the $tr(2)$ term of eq. (7 c).
 Let us focus in the following on the "++" terms
(the others being obtained by symmetry) and extract
 the potentially singular part from the
$\tau_+ \tau'_+$ term. The expression of $\tau_+$ (cf. eq. (10 c))
shows that terms in $\alpha_+ \ \beta '_+$, $\alpha '_+ \ \beta_+$, $\alpha_+
\ \alpha '_+$ and $\beta_+ \ \beta '_+$ will appear in the singular
contribution,
because each of the latter quantities can be associated with a ${\wpp} \left (
{1 \over Qk} \right ) \ \delta (Qk')$ product.\par
The contribution is~:
\vskip 2 mm
$$\left . Im \ \Pi_{\mu}^{\mu} \right |_{2, reg, + +} = 2 m_f^4  \int~[dk ]
\int {d \Omega_1 \over 4 \pi}
 {\vert_{reg}} \int { d \Omega_2 \over 4 \pi} {\vert_{reg}}~(Q_1Q_2) \
\eqno(17)$$
$$ \cdot \left [ \left ( Q_1 k_+ \right ) \left ( Q_2 k'_+ \right ) + \left (
Q_1 k'_+ \right )
\left ( Q_2 k_+ \right ) - \left ( Q_1 \ Q_2 \right ) \left ( k_+ k'_+ \right )
\right ]$$
$$ \cdot~ \left \{ {1 \over (Q_1 P)} {\wpp} \left ( {1 \over Q_2 k'} \right )
{\wpp}
\left ( {1 \over Q_2 k} \right ) \delta \left ( Q_1 k \right ) \alpha_+
\beta '_+ \right .$$
$$+ {1 \over (Q_1 P)} {\wpp} \left ( { 1 \over Q_2 k} \right )
{\wpp} \left ( {1 \over Q_2 k'} \right ) \delta \left ( Q_1 k' \right )
\alpha '_+ \beta_+ $$
$$\left. - {1 \over (Q_1 P)} {1 \over (Q_2 P)} \left ( \delta \left ( Q_1 k
\right ) \delta \left ( Q_2 k' \right ) \alpha_+ \alpha '_+ - \pi^2 \delta
\left
(Q_1 k' \right ) \delta \left ( Q_2 k' \right ) \beta_+ \beta '_+ \right )
\right \} \ \ \ .$$
\vskip 3 mm

 We find the ${1 \over
\hat{\varepsilon}}$ contribution for
 $\widehat{Q_1} \to - {\vec{p} \over E}$ or $\widehat{Q_2} \to - {\vec{p}
\over E}$ (the latter is obtained by symmetry $k \leftrightarrow k'$).
 No double pole has to be considered because of the presence of the
 $(Q_1 Q_2)$ factor.
 \par
Using ~:
\vskip 2 mm
$$\int {d \Omega \over 4 \pi} \delta (Qk) =
 {\theta (- k^2) \over 2 | \vec{k}|} \ \ , \eqno(18)$$
$$\int {d \Omega \over 4 \pi} \delta (Qk') (Qk) = {\theta (- k'^2) \over 2
|\vec{k}'|^2} \left (
\omega |\vec{k}'| - \omega ' \vec{k} \cdot \widehat{k'} \right ) \ \ ,$$ \vskip
3 mm
\noindent and [11]~:
$$\int {d \Omega \over 4 \pi} {\wpp} \left ( {1 \over kQ} \right ) = L(k)
\equiv L = {1
\over 2k} \ell n \left | {\omega + k \over \omega - k } \right | \ \ ,
\eqno(19)$$
$$\int {d \Omega \over 4 \pi} {\wpp} \left ( {k'Q \over kQ} \right ) = \omega '
L -
{\vec{k} \cdot \vec{k'} \over \vec{k}^2}  \left ( \omega L - 1 \right ) \ \ \
,$$
\vskip 3 mm
\noindent we obtain after combining the contributing terms~:
$$ Im \ \Pi_{\mu}^{\mu} \vert_{2, reg, + + } =
 \ m_f^4  ~
 {1 \over \widehat{\varepsilon}}
 \int~[dk]~ {\delta (Pk) \over E} \ \ \ \ \eqno(20)$$
 $$  \cdot {\lbrace }
\left ( \alpha_+ \beta '_+  + \alpha '_+ \beta_+ \right )
 \ \left [ 2 \left ( 1 - \omega / k \right ) \left ( 1 - \omega ' / k'
\right )  L
          + \left (  \left ( E - k \right ) / k'  - 1 \right )
 \left ( \left ( 1 - {\omega^2
\over \vec{k}^2} \right ) L + {\omega \over \vec{k}^2} \right ) \right ]  $$
$$  +~  \left ( -
\alpha_+ \alpha '_+ + \pi^2 \beta_+ \beta '_+ \right )
{}~ {1 \over 2k} \left [ 2 \left ( 1 - \omega /k \right ) \left
( 1 - \omega ' /k' \right ) + \left ( 1 - {\omega^2 \over k^2} \right )
\left (  \left ( E - k \right ) / k' -   1 \right )
 \right ] $$
  $$ +~ sym \left ( \omega, k
\longleftrightarrow \omega' , k' \right ) {\rbrace } \  \  . $$
\vskip 3mm

\noindent Similar contributions come
 from $\tau_+ \tau '_-$ and  $\tau_- \tau '_-$, respectively.
\par
\vskip 2 mm
\noindent The functions $L$ (and $L'$) in eq. (20)
 are eliminated by using the
definitions for $\alpha_{\pm}$ and $\beta_{\pm}$, eq. (10 d)~:
\vskip 2 mm
$$m_f^2 \left ( 1 - \omega /k \right ) L \ \alpha_+ = \left ( \omega - k -
{m_f^2 \over
k} \right ) \alpha_+ - \pi^2 {m_f^2 \over 2k} \left ( 1 - \omega /k \right )
\beta_+
+ 1 \ \ , \eqno(21)$$
$$m_f^2 \left ( 1 - \omega /k \right ) L \ \beta_+ = \left ( \omega - k -
{m_f^2 \over
k} \right ) \beta_+ + {m_f^2 \over 2k} \left ( 1 - \omega /k \right ) \alpha_+
\ \
\ . $$ \vskip 3 mm

 \noindent Thus eq. (20) contains  terms proportional to $\alpha_+ \alpha
'_+$, $\alpha_+ \beta '_+$,
 $\alpha '_+ \beta_+$
 and $\beta_+ \beta '_+$, and  terms linear in $\beta_+$, $\beta
'_+$.
The $\alpha_+ \alpha '_+$ and $\beta_+ \beta '_+$ terms vanish in
 eq. (20), whereas the $\alpha_+ \beta '_+$ and $\alpha '_+ \beta_+$ terms
compensate
with those of eq. (16). All what remains are the linear terms in $\beta_+ $ and
$\beta
'_+$. The final result (including all contributions from $\tau_{\pm} \tau
'_{\pm}$) is~:
\vskip 2 mm
$$\left . Im \ \Pi_{\mu}^{\mu} \right |_{reg} = 2 \ m_f^2  {1 \over
\widehat{\varepsilon}} \int~[dk]~ {\delta ( Pk )\over E} \   \eqno(22)$$
$$\cdot \ \left \{ \beta '_+ \left ( 1 - \omega '/k' \right ) + \beta_+ \left (
1 - \omega/k
\right ) + \beta '_- \left ( 1 + {\omega ' \over k'} \right ) + \beta_- \left (
1 +
{\omega \over k} \right ) \right \} \ \ \ .$$
\vskip 3mm
This expression may still be simplified  by following a procedure
familiar from the hard photon case [6].
Since the functions $\beta_+$ and $\beta_-$ as given by [13 - 17]
 are peaked for $\omega \to 0$ ($k$ being fixed)  we may
replace $n_F(\omega) \ n_F (E - \omega)$ by $n_F(0) \ n_F(E)
= {1 \over 2} n_F (E)$. After performing the angular integration in
eq. (22), the remaining integrals $\int d \omega (1 \mp {\omega \over k} )
\beta_{\pm}( \omega , k)$ are evaluated using the sum rule [6]~:
\vskip 2 mm
$$\int_{- \infty}^{\infty} d \omega \left ( 1 \mp \omega /k \right ) \rho_{\pm}
( \omega
, k) = 0 \ \ . \eqno(23)$$
\vskip 3 mm
\noindent The dominant contribution of the integral over $k$
 comes from $m_f < k < T$.
The leading contribution for $g \to 0$ then reads :
\vskip 2 mm
$$E {dW \over d^3 \vec{p}} \simeq {1 \over \widehat{\varepsilon}} \ { e_q^2 \
\alpha \alpha_s
\over 2 \pi^2}~ T^2 \ n_F(E) \left ( {m_f \over E} \right )^2 \ell n \left ( 1
\over {\alpha_s} \right ) \ \ \ . \eqno(24)$$
\vskip 3 mm
This result shows that the Braaten-Pisarski resummation does not
 yield a finite soft real photon production rate:
a logarithmic divergence remains.
 \par
\vskip 5 mm

\section{4}{DISCUSSION}

The above analysis allows to identify the diagrams which are responsible
for the singularities as they originate  from terms proportional
to the product ${\wpp} \left ( {1 \over Qk} \right )
\delta (Qk') $.
One example of such a diagram is shown in Fig. 2, where the singularity
 is due to the massless quark exchange present in the hard thermal loop
 effective vertices. The massless exchange is transparent in
 the two $\rightarrow$ three amplitude of Fig. 2b.
The singularity arises from the configuration $Q\cdot P = 0$: it corresponds
to a collinear singularity when
 the photon is allowed to stay massless, $P^2 = 0$. \par
 At present we do not know how to screen this mass singularity
by a consistent prescription. Therefore a pragmatic approach
is to introduce a soft cut-off of $O( gT )$ in order to regularize
 the soft photon production rate at logarithmic accuracy~:
\vskip 2 mm
$$E {dW \over d^3 \vec{p}} \simeq { e_q^2 \ \alpha \alpha_s }~ T^2 \
\ell n^2 \left ( {1 \over \alpha_s} \right ) \ \ \ , \eqno(25)$$
\vskip 3 mm
\noindent where the photon energy is assumed $E \sim m_f$. \par

The presented result is valid for soft massless, i.e. non-thermalized photons.
This implies that the quark - gluon plasma has to have a finite size;
its characteristic length is denoted by $L$.
As already mentioned in the Introduction the mean free path $l_{\gamma}$
of the detected photon has to be larger than $L$.
Since only photons with wave lengths less than $L$ are radiated,
the dimension of the plasma becomes constrained [18] ~:
\vskip 2 mm
$$  { 2 \pi \over E} < L  < l_{\gamma} (E)
\ \  .  \eqno(26)$$ \vskip 3 mm
\noindent Suppressing the logarithmic factors in eq. (25)
we estimate the photon's mean free path to be given by~:
\vskip 2 mm
$$ l_{\gamma} \simeq { E \over { \alpha \alpha_s T^2}} \ . \eqno(27)$$
\vskip 3 mm
\noindent This order of magnitude estimate is in agreement
 with $ l_{\gamma} \simeq { 1 / n_q \sigma_{Compton} (E)} $,
where $n_q$ is the quark density and
  $  \sigma_{Compton} (E) \sim \alpha \alpha_s / E T $ is the
  high energy Compton cross section in the QGP, which
is responsible for the photon absorption.
  \par
For soft photon energies $E \sim m_f \sim O(gT)$ the constraint eq. (26)
becomes~:
\vskip 2 mm
$$ O ( 1 ) ~ < ~ L T \sqrt{\alpha_s}~ < ~ O( {1 \over \alpha })
\ \  ,  \eqno(28)$$ \vskip 3 mm
\noindent i.e. for typical values of $T \sim  400 MeV$ and
$\alpha_s \sim 0.25$ the constraint reads :
$ 1 fm < L < 100 fm$. This size is compatible with the expectations for
a realistic QGP produced in heavy - ion collisions. \par

In summary it seems reasonable to forsee experimental  situations
where soft $O(gT)$ non-thermalized photons would be emitted from
a QGP. However, our present understanding does not allow us
to derive their finite production rate.
\vfill \supereject

 \par\vskip 15pt
\noindent{\bf ACKNOWLEDGEMENTS}\par
We kindly thank  P.~Aurenche, J.~Kapusta and
   R.~D.~Pisarski
 for helpful remarks, and E.~Pilon for discussions.
Partial support of this work by
 "Projets de Coop\'eration et d'Echange"
(PROCOPE) is gratefully acknowledged.
\vskip 1.00 truecm
\noindent{\bf REFERENCES}\par
\baselineskip=15pt

\item{1)}
 $\,$ For a recent review: P.~V.~Ruuskanen, in
{\it Particle Production in Highly Excited Matter},
eds.  H.~H.~Gutbrod and J.~Rafelski,
Proc. of ASI, Il Ciocco, Lucca (Italy), 1993

\item{2)}
 $\,$ K.~Kajantie and H.~I.~Miettinen,
Z. Phys. \cit {C9} {1981} {341};
and earlier references quoted therein.

\item{3)}
 $\,$ K.~Kajantie and P.~V.~Ruuskanen,
Phys. Lett. \cit {B121}  {1983} {352}.

\item{4)}
 $\,$ M.~Neubert,
Z. Phys. C - Particles and Fields  \cit {C42} {1989} {231}.

\item{5)} $\,$
 J.~I.~Kapusta, P.~Lichard, and D.~Seibert,
 Phys. Rev.  {\bf D44} (1991) 2774.

\item{6)} $\,$
 R.~Baier, H.~Nakkagawa, A.~Ni\'egawa, and K.~Redlich,
 Z. Phys. C - Particles and Fields  {\bf C53} (1992) 433.

\item{7)}  $\,$
E.~Shuryak and L.~Xiong,
 Phys. Rev. Lett. {\bf 70} (1993) 2241;
 A.~Makhlin, preprint SUNY-NTG-93-10, January 1993.

\item{8)} $\,$
 For a review and references: R.~D.~ Pisarski, Nucl.
Phys. {\bf A525} (1991) 175c;  and E.~Braaten, Nucl. Phys.(Proc.
Suppl.) {\bf B23} (1991) 351.

\item{9)} $\,$
 R.~D.~Pisarski, Nucl. Phys. {\bf B309} (1988) 476;
 Phys. Rev. Lett. {\bf 63} (1989) 1129.

\item{10)} $\,$
 E.~Braaten and R.~D.~Pisarski, Phys. Rev. Lett. {\bf
64} (1989) 1338; Nucl. Phys.  {\bf B337} (1990) 569; Nucl.
Phys. {\bf B339} (1990) 310.

\item{11)} $\,$
 J.~Frenkel and J.~C.~ Taylor, Nucl. Phys. {\bf B334}
(1990) 199.

\item{12)}  $\,$
 T.~Altherr and P.~V.~Ruuskanen,
Nucl. Phys. {\bf B380} (1992) 377.

\item{13)} $\,$
E.~Braaten, R.~D.~Pisarski, and T.~C.~Yuan, Phys. Rev. Lett.
 {\bf 64} (1990) 2242;

\item{14)} $\,$
 S.~M.~H.~ Wong,
 Z. Phys. C - Particles and Fields  {\bf C53} (1992) 465.

\item{15)} $\,$
 V.~V.~Klimov, Sov. J. Nucl. Phys. {\bf 33} (1981) 934;
O.~K.~Kalashnikov, Fortschr. Phys. {\bf 32} (1984) 525.

\item{16)} $\,$
H.~A.~Weldon, Phys. Rev. {\bf D26}  (1982) 2789;
  Physica \cit {A158} {1989} {169}; and
 Phys. Rev. \cit {D40} {1989} {2410}.

\item{17)} $\,$
 R.~D.~Pisarski, Physica {\bf A158} (1989) 146;
Fermilab preprint Pub - 88/113-T (unpublished).

\item{18)} $\,$
For a discussion of soft photon radiation in a QED plasma:
H.~A.~Weldon, Phys. Rev. {\bf D44}  (1991) 3955.
\vskip 0.50 truecm
\vfill\eject
\vskip 0.50 truecm
\noindent{\bf FIGURE CAPTIONS}\par
\medskip
\item{Fig.1:}
One-loop diagram for the production of a real soft photon (weavy line)
with momentum $P$. The effective quark propagator and the
effective quark-photon vertex are indicated by a blob.
\item{Fig.2:} (a) Cutting the effective one-loop diagram  through
 the effective hard thermal loop vertex gives rise to
 (b) the amplitude with a collinear singularity for $P \cdot Q = 0$.
 The curly line denotes the gluon.

\bye